\documentclass{paper} 
\usepackage{color}

\usepackage{amsmath}
\usepackage{amsfonts}
\usepackage{amssymb}

\usepackage{epsf}
\usepackage{graphicx}
\usepackage{gensymb}
\usepackage{float}

\def\beq#1{\begin{equation}\label{#1}}
\def\eeq{\end{equation}}
\def\beqa#1{\begin{eqnarray}\label{#1}}
\def\eeqa{\end{eqnarray}}

\def\myfrac#1#2{\left(\frac{#1}{#2}\right)}
\def\comment#1{\relax}

\def\apgt{\ {\raise-.5ex\hbox{$\buildrel>\over\sim$}}\ }
\def\aplt{\ {\raise-.5ex\hbox{$\buildrel<\over\sim$}}\ }

\begin{document}

\title{Possible electromagnetic signatures of coalescing neutron -- black hole binaries}

\author{
K.A. Postnov$^{1,2,3}$,
A.G. Kuranov$^{1,4}$,
I.V. Simkin$^{5}$}
  
  \institution{$^1$  Sternberg Astronomical Institute, Universitetskij pr. 13,\\ 119234 Moscow, Russia\\
  $^2$ Institute of Experimental and Theoretical Physics,\\
  B. Cheremushkinskaya 25, 
  117218 Moscow, Russia\\
  $^3$ Faculty of Physics, Novosibirsk University, 
  Pirogova 2, \\
  630090 Novosibirsk, Russia\\
  $^4$ Foreign Trade Academy, Pudovkina 4a,\\ 119285 Moscow, Russia\\
  $^5$ Baumann Moscow Technical University,
  2d Baumanskaya 5, \\105005 Moscow, Russia}

\maketitle

\begin{abstract}

Possible models for the generation of electromagnetic (EM) radiation during the coalescence of neutron star–black hole binaries are considered. The mass of the remnant disk around the black hole during the coalescence of neutron stars and black holes is calculated by taking into 
account the equation of state for neutron stars and the rotation of the binary components before the coalescence. The parameters of binary systems before the coalescence (the mass ratio, the 
component rotation, the neutron star magnetic field) are calculated by the population synthesis method. The derived mass of the remnant disk around the black hole after the coalescence is used to 
estimate the kinetic energy of the relativistic jet launched by the Blandford–Znajek mechanism. A disk mass of more than $\sim 0.05 M_\odot$ required for the formation of short gamma-ray bursts is shown to be obtained in no more than 1-10\% of the 
coalescences (depending on the equation of state). Less efficient common envelopes (a large parameter $\alpha_{CE}$) lead to a noticeably larger percentage of events with astrophysically interesting EM energy release. For binaries with a large mass ratio, in which a magnetized neutron star is not subjected to tidal disruption before the coalescence, the possibility of the formation of an electrically charged rotating black hole (Wald charge) is considered and estimates of the maximum EM power released by such a black hole after the coalescence are made. The conversion of the emitted gravitational waves into electromagnetic ones in the relativistic lepton plasma generated in coalescing pulsar–black hole binaries at the pre-coalescence stage is also discussed.

\keywords{gravitational waves, neutron star -- black hole binaries, gamma-ray bursts}

\end{abstract}

\section{INTRODUCTION}
\label{s:intro}

The discovery of gravitational waves (GWs) from the coalescence of binary black holes (BHs) by the LIGO ground-based laser interferometers (Abbott et al. 
2016b) was one of the biggest scientific discoveries at the beginning of the 21st century and gave powerful impetus to the development of 
multi-wavelength observations of cosmic transients. Multi-wavelength observations of the first binary neutron star (NS) coalescence GW170817 (Abbott et al. 2016a) ushered in a new era of astronomical observations -- multi-messenger astronomy. At present, there is detailed information about ten coalescing binary BHs discovered by the LIGO/Virgo Collaboration during the first and second scientific run O1 and O2 (LIGO/Virgo Collaboration 2018). More than 30 coalescing binary BHs, several NS + NS binary candidates, and several NS + BH binary candidates were discovered in the ongoing O3 observations by the LVC Collaboration (see the online catalog https://gracedb.ligo.org/latest/). 
No reliable detection of the accompanying electromagnetic (EM) radiation from new coalescences in the O3 data has been reported so far. Obviously, the information about the GW coalescence source obtained from EM observations complement significantly the information obtained from an analysis of the GW signal properties. For example, the latest analysis of the GW data by the LVC Collaboration (LIGO Collaboration et al. 2019) does not rule out the possibility that one of the GW170817 binary components can be a low-mass BH, although an analysis of the EM radiation from the accompanying gamma-ray burst GRB 170817A argues for the formation of a supermassive neutron star as a result of the coalescence and, hence, for the model of two coalescing NSs in the source GW170817 (Gill et al. 2019; Piro et al. 2019).

Clearly, studying the generation mechanisms and parameters of the EM radiation during the coalescence of binary relativistic stars remains a topical problem. In this paper, we address NS + BH binary systems. Such binaries are of interest on 
their own because a magnetized NS before the coalescence can be a radio pulsar. An analysis of the pulse arrival time (timing) for such a pulsar in a strong BH gravitational field could provide unique information about the spacetime structure near the BH. Binary radio pulsars with BHs have been studied previously (see, e.g., Lipunov et al. 1994; Pfahl et al. 2005), but they have not yet been detected.

As relativistic numerical calculations show, the result of the NS–BH coalescence in a binary system depends significantly on the component mass ratio $q=M_{BH}/M_{NS}>1$ and the NS equation of state (Shibata and Taniguchi 2011; Shibata and Hotokezaka 2019). The neutron star can be disrupted by tidal forces before the coalescence or be swallowed by the BH entirely. A key criterion is the ratio of
the tidal NS radius $R_t\sim R_{NS}q^{1/3}$ to the radius of the innermost stable circular orbit around the BH $R_{ISCO}$. The tidal radius depends on the mass ratio and the NS equation of state, while the radius of the innermost stable circular orbit is determined by the BH mass and angular momentum. At $R_t>R_{ISCO}$ a disk-shaped or crescent-shaped structure is formed around the BH with a possible sub-relativistic dynamic jet as a result of the coalescence (Kyutoku et al. 2015; Shibata and Hotokezaka 2019), which is favorable for the emergence of the subsequent optical kilonova afterglow (Kawaguchi et al. 2016; Metzger 2019). After the coalescence, the BH acquires an additional angular momentum and physical conditions arise for the formation of a relativistic jet, for example, by the Blandford–Znajek (BZ) EM mechanism (Blandford and Znajek 1977), and the generation of a short gamma-ray burst (GRB) (Nakar 2007).

In our recent paper (Postnov and Kuranov 2019), we performed model calculations of the angular momenta of coalescing binary BHs for various initial spins of the components (co-aligned and randomly oriented spins), BH formation models (without any additional fallback of the stellar envelope during the collapse onto the CO core and with this fallback), and various common envelope efficiencies $\alpha_{CE}$  (the ratio of the binding energy of the stellar core and envelope after the main sequence to the orbital energy of the binary before the beginning of the common envelope stage) by the population synthesis method.
In these calculations we used the standard scenario
for the evolution of massive binary stars (Postnov and
Yungelson 2014) supplemented by the treatment of
the evolution of stellar core rotation with allowance
made for the effective core–envelope coupling proposed
in Postnov et al. (2016). The calculations were
performed for various initial chemical compositions
(metallicities) of stars by taking into account the time
evolution of the metallicity and the star formation
rate in the Universe (for details, see Postnov and
Kuranov (2019)). The technique of these calculations
was applied to NS+BH binaries, which
allowed the coalescence rate ${\cal R}$ (in yr$^{-1}$ Gpc$^{-3}$) and the detection rate 
${\cal D}$ (in yr$^{-1}$) at the sensitivity level of the operating GW interferometers to be calculated (Postnov et al. 2019).

In the present paper, the mass ratios $q=M_{BH}/M_{NS}$ and
the BH spins before the coalescence of NS+BH
binaries obtained in our calculations (Postnov
et al. 2019) are used as input parameters to determine
the mass of the remnant disk around the BH $M_{d}$
and the BH spin (characterized by the dimensionless
Kerr parameter $a^*=J_{BH}/(GM_{BH}^2/c^2)$, $J_{BH}$ 
is the BH angular momentum, $G$ is the gravitational
constant, and $c$ is the speed of light) after
the coalescence. The disk mass depends significantly
on the NS compactness (mass-to-radius
ratio $M_{NS}/R_{NS}$), which is defined by the NS equation
of state. The effects of the equation of state
are parameterized by the tidal deformability
$\Lambda=(2/3)k_2[(c^2/G)(R_{NS}/M_{NS})]^5$
(Damour et al. 2012)
($k_2$ is the Love number). This parameter is constrained
from the GW observations of the source
GW170817 (Abbott et al. 2019).

In turn, the disk mass and the BH spin determine
the possible kinetic energy of the relativistic jet
launched by the BZ mechanism. The kinetic energy
of the jet $\Delta E_{BZ}$ may be considered as an upper limit
for the isotropic energy of a short GRB $\Delta E_{iso}$. For
coalescing binaries with a large mass ratio $q$, in which
the NS is swallowed by the BH without disruption,
the NS magnetic field and the BH spin after the
coalescence $a^*$ are used to calculate the possible BH
electric charge in the NS magnetic field (Wald 1974).
In addition, if the NS before the coalescence was at
the radio pulsar stage, then the medium around the
coalescing NS+BH binary could be filled with a
relativistic lepton plasma. We also consider the mechanism
for the conversion of gravitational waves into
electromagnetic ones in such a magnetized relativistic
plasma.

\section{COALESCENCE AND DETECTION RATES
OF NS+BH BINARIES WITH
GW INTERFEROMETERS}
\label{s:rates}

Figure 1 shows the results of our population synthesis
calculations of the NS+BH binary coalescence
rate density and the detection rates by the
LIGO/Virgo ground-based laser GW interferometers.
The parameters of the NS and BH formation, the
evolution of massive binary systems, and the technique
of calculations are described in detail in Postnov
and Kuranov (2019) and Postnov et al. (2019).
At the coalescence phase, the GW signal amplitude is
determined by the chirp mass of the binary
system, which for two point masses $M_1$ and $M_2$
is ${\cal M}=(M_1M_2)^{3/5}/(M_1+M_2)^{1/5}$. The detection
horizon $D_h$ of coalescing binary systems with a chirp
mass  ${\cal M}=1.22 M_\odot$ for two neutron stars with a
canonical mass of 1.4 $M_\odot$ in the current LIGO/Virgo
observations is taken to be 120 Mpc. The detection
horizon depends on the binary chirp mass as $D_h\sim {\cal M}^{5/6}$ (LIGO Collaboration et al. 2010).

The distribution of parameters for binaries that can
be recorded by the ground-based GW interferometers
within the detection horizon $D_h$ is important from
the viewpoint of comparison with observations. At
a given sensitivity level of the GW detectors, $D_h$ is
determined mainly by the chirp mass. Thus, for each
chirp mass there is a limiting distance (redshift) to
which the binary coalescence rate density ${\cal R}_{NSBH}$
calculated by the population synthesis method should
be integrated. In turn, the coalescence rate per unit
comoving volume ${\cal R}_{NSBH}(t)$ is a convolution of the
calculated time dependence of the binary system coalescence
rate ${\cal R}_\delta(t)$ calculated for an instantaneous
star formation burst with the star formation rate and
the stellar metallicity as functions of time $SFR(t)$:
${\cal R}_X(t)=\int^t SFR(t-\tau){\cal R}_\delta(\tau)d\tau$. 
The time evolution of the metallicity of the stellar population and the star formation rate were calculated using the formulas given in Postnov and Kuranov (2019).

It can be seen from Fig. 1 that the expected detection rate of NS+BH binaries is approximately several events per year, which is consistent with the existing detection statistics of such binaries (for the online information about the detected events, see https://gracedb.ligo.org/latest/).

\begin{figure}
\includegraphics[width=0.65\columnwidth]{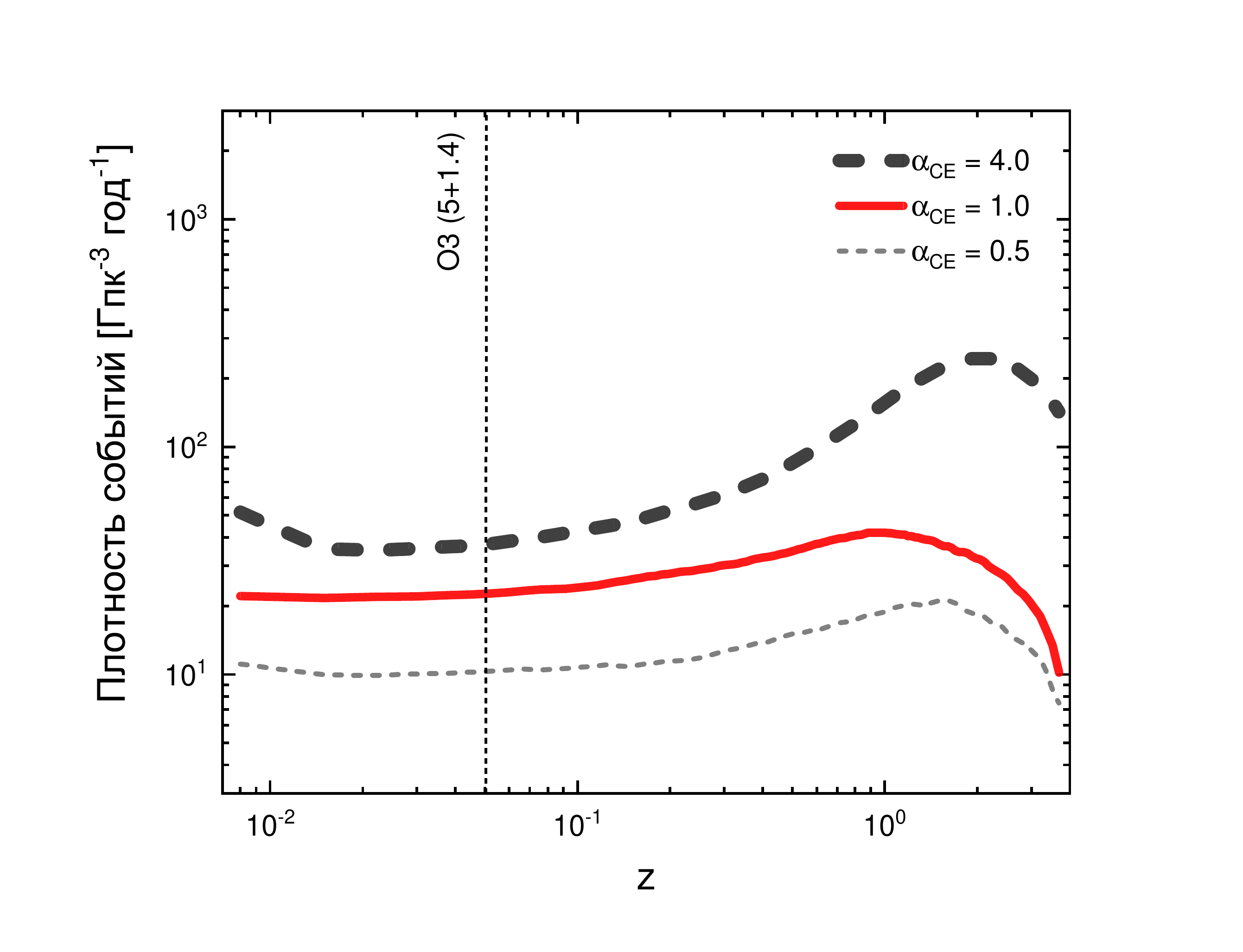}
\includegraphics[width=0.65\columnwidth]{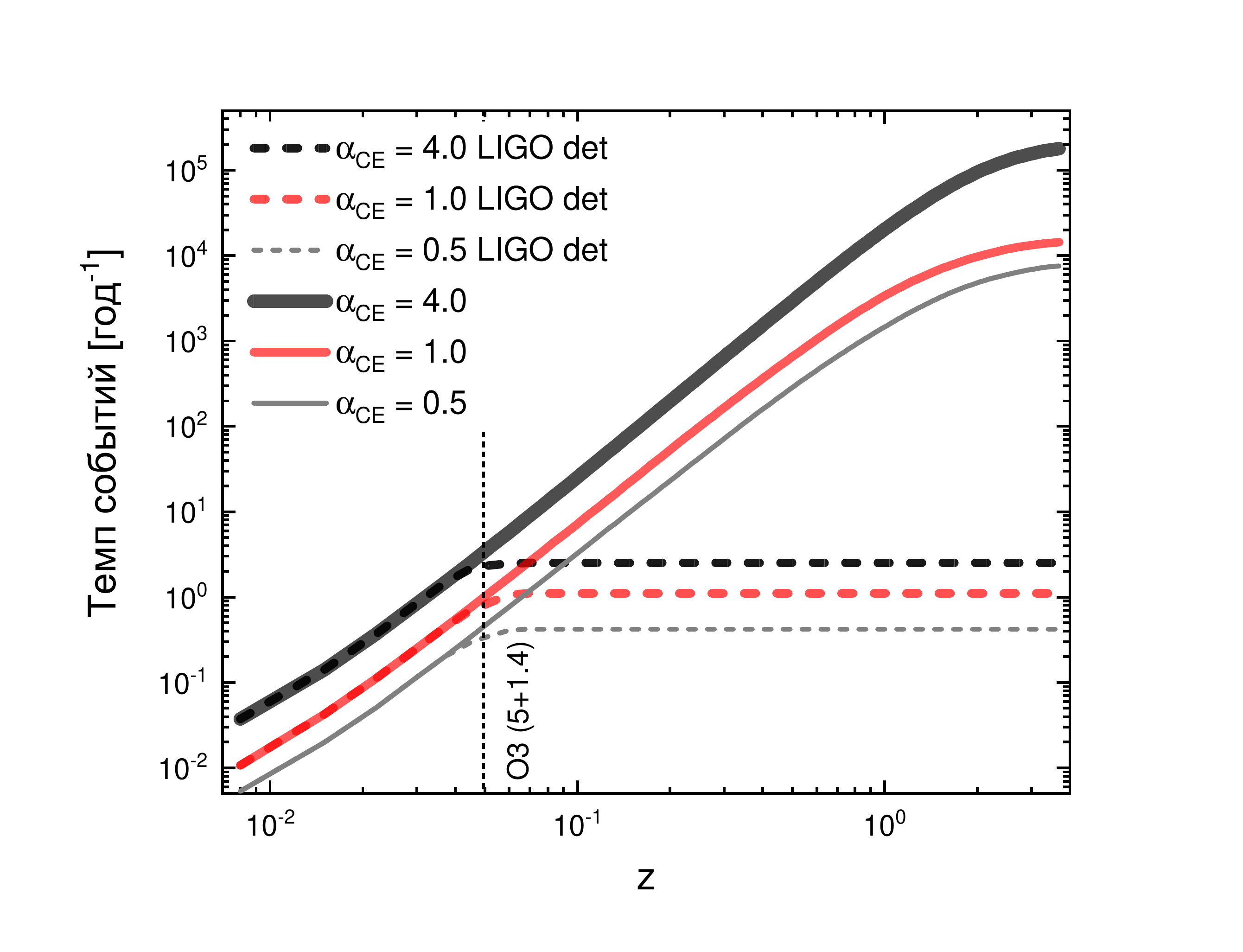}
\caption{
Upper panel: NS+BH coalescence rate density ${\cal R}_{NSBH}$ (per year per cubic Gpc) versus cosmological redshift $z$ for various values of the parameter $\alpha_\mathrm{CE}$  (common envelope efficiency) with allowance made for the evolution of the mean star formation rate and the stellar metallicity in the Universe. The upper and lower boundaries (dashed lines) correspond to $\alpha_\mathrm{CE}= 4$ and 0.5, respectively; the solid line corresponds to $\alpha_\mathrm{CE}= 4$. The vertical dashed line indicates the LIGO/Virgo O3 detection horizon for coalescing binary systems with masses of 5+1.4 $M_\odot$. Bottom panel: Number of NS+BH coalescences per year (the integral of the coalescence rate per unit volume to the distance corresponding to a given $z$) versus limiting redshift (detection horizon $D_h$) with allowance made for the star formation history in the Universe for common envelope efficiencies $\alpha_\mathrm{CE}=0.5,\, 1,\, 4$. The dashed curve indicates the expected number of events per year detected by the LIGO/Virgo O3 interferometers for the averaged orientation of the binary orbits relative to the line of sight 
${\cal R}_\mathrm{BHNS}\sim 1-3$ yr$^{-1}$.The vertical dashed line indicates the LIGO/Virgo O3 detection horizon 
for coalescing binary systems with masses of 5+1.4 $M_\odot$.
}
\end{figure}

\section{COALESCENCES WITH NS TIDAL DISRUPTION}
\label{s:tidal}

The NS+BH binary coalescences in which the NS is tidally disrupted are most interesting from the standpoint of observational manifestations in the EM range. As was said in the Introduction, the NS tidal disruption occurs mostly in binaries with a small
component mass ratio  $q=M_{BH}/M_{NS}\lesssim 3$ and depends on the NS equation of state (tidal deformability $\Lambda$).

\subsection{Remnant Disks around the BH}
\label{ss:disk}

To estimate the mass of the baryonic disk (torus) around the BH left after the coalescence, we will use the fit to the numerical data with allowance made for the NS equation of state that has been proposed recently (Zappa et al. 2019). In turn, the
fitting formulas in this paper use the results from Jimenez-Forteza et al. (2017), where the radiated GW energy, the mass and spin of the BH remaining after the coalescence of binary BHs with a different mass ratio are calculated.

The derived mass ratio distributions of NS+BH binaries are presented in Fig. 2. The upper three rows in this and next figures present the simulation results for stellar metallicities in the ranges 
 $Z>0.01$, $0.01Z>0.001$ and $0.001Z>0.0001$ 
(from top to bottom), while the lower row presents the convolution with the time evolution of the metallicity. 

The left and right columns in the figure present the calculations, respectively, for the standard common envelope efficiency $\alpha_{CE}=1$ and $\alpha_{CE}=4$
corresponding to a smaller approach of the binary components in the common envelope. Note that the reduced common envelope efficiency $\alpha_{CE}=4$ corresponds better to the treatment of the binary dynamics in the common envelope based on the angular momentum conservation law (the so-called $\gamma$-formalism; see Nelemans and Tout 2005) and is required to explain the properties of the population of symbiotic X-ray binary systems in the Galaxy (Yungelson et al. 2019). It can be seen from Fig. 2 that in $\sim 10-20\%$ of the coalescences the component mass ratio is small enough for the formation of a remnant disk around the BH after the coalescence.

The BH spin after the coalescence of NS+BH binaries is entirely determined by the initial BH spin
$a^*$ and the mass ratio $q$ and depends weakly on the
NS equation of state. The distribution in initial BH
spins (before the coalescence) calculated by Postnov
et al. (2019) is indicated in Fig. 3 by the dotted line,
while the spins after the coalescence $a^*_f$ are indicated
by the solid line. The BH spins after the coalescence
are seen to be concentrated near $a^*_f\sim 0.5$, while the fraction of rapidly rotating BHs after the coalescence is small.

The resulting mass of the disk around the BH after
NS tidal disruption is shown in Fig. 4 for various
values of the NS tidal deformability $\Lambda$ in the wide
range from 100 to 2000 spanning a broad spectrum
of possible NS equation of state (LIGO Collaboration
et al. 2019). We see a strong dependence of the disk
mass on the NS equation of state -- astrophysically
interesting disk masses $M_d > 0.05 M_\odot$ are obtained
only at large $\Lambda > 300$ corresponding to small compactness $M_{NS}/R_{NS}$ (stiff equations of state). Note
that the constraints on the parameter $\Lambda$ from the GW
observations of GW170817 lie within a wide range,
but significant tidal deformations with $\Lambda\gtrsim 1600$ are
highly unlikely (Abbott et al. 2019; LIGO Collaboration
et al. 2019). An independent analysis with the
involvement of other constraints gives $\Lambda = 390^{+280}_{-210}$
for the mass $M_{NS} = 1.4 M_\odot$ (Jiang et al. 2019).  

\begin{figure*}
	\includegraphics[width= \columnwidth]{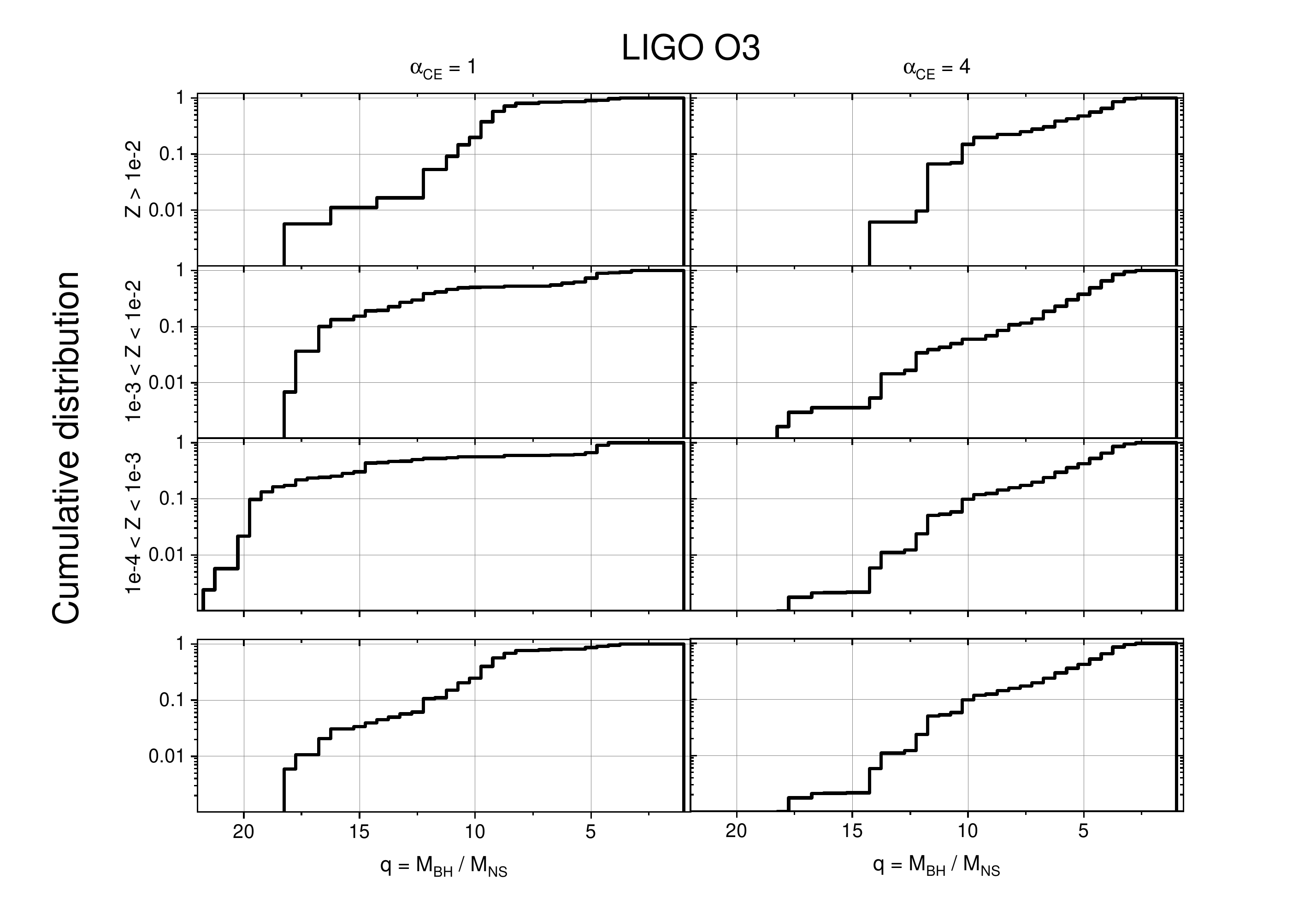}
	\caption{Cumulative distribution in mass ratio in coalescing NS+BH binaries that can be recorded at the sensitivity level of the LIGO/Virgo GW detectors in the observing run O3. The upper three rows present the results for various stellar metallicities. The fourth row presents the result with allowance made for the time evolution of the mean star formation rate
and the stellar metallicity in galaxies. The left and right columns present the calculations for the common envelope parameters $\alpha_{CE}=1$ and 4, respectively.	}
	\label{fig:qO3}
\end{figure*}

\begin{figure*}
	\includegraphics[width=\columnwidth]{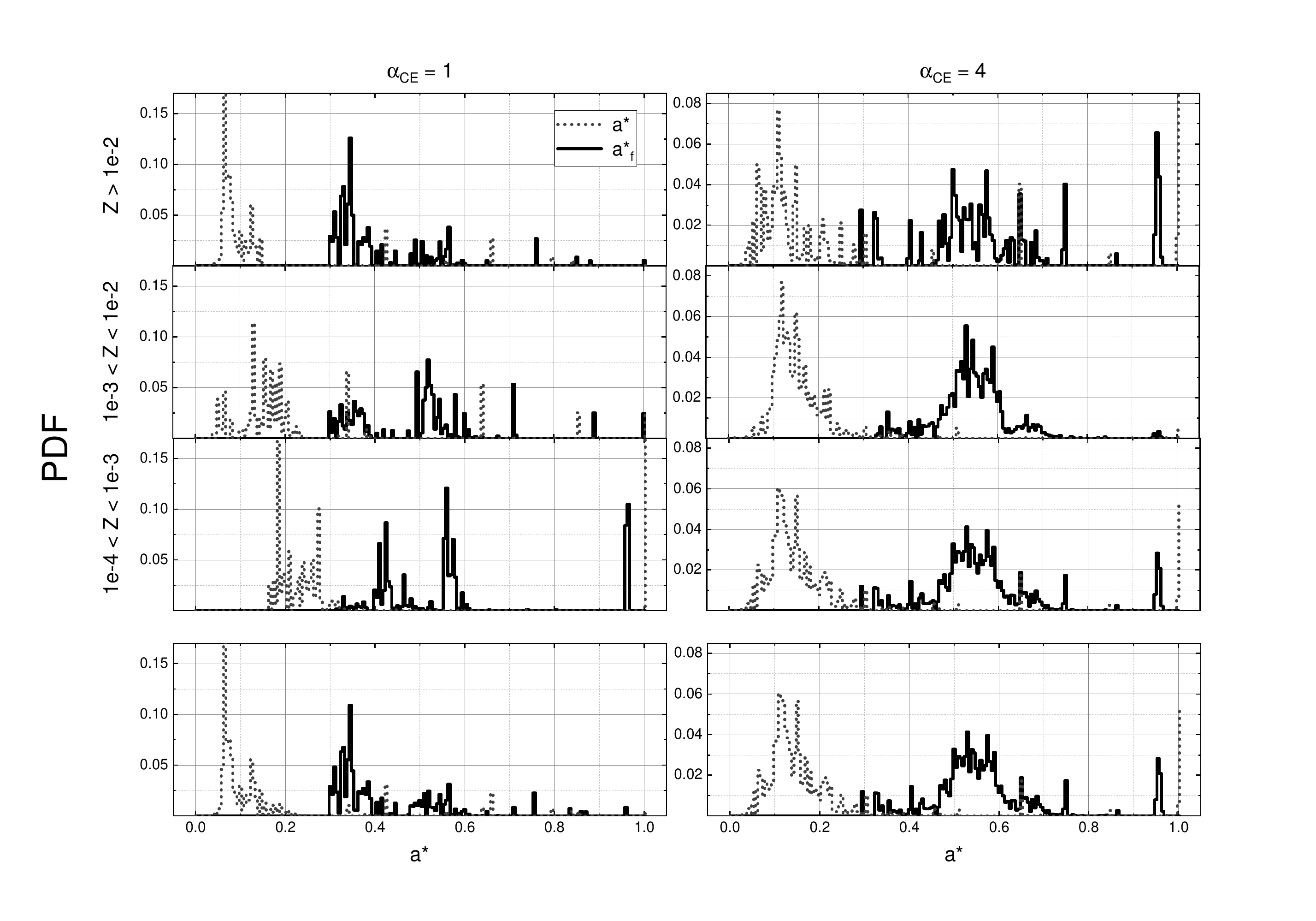}
	\caption{BH spins before ( $a^*$, dotted line) and after the coalescence ($a^*_f$, solid line) in NS+BH binaries. The dependence on the NS equation of state is indistinguishable. The upper three rows present the results for various stellar metallicities $Z$. The fourth row presents the convolution with allowance made for the time evolution of the mean star formation rate and the stellar metallicity in galaxies. The left and right columns present the calculations for the common envelope parameters $\alpha_{CE} = 1$ and 4, respectively.}
	\label{fig:spins}
\end{figure*}

\subsection{Jet kinetic energy}

The minimum isotropic kinetic energy of the
relativistic jet required to produce a GRB is estimated
from observations as $\Delta E_{K,min}\simeq 10^{48}$ erg (Soderberg
et al. 2006). An analysis of the observations of short
GRBs shows that the mean conversion efficiency of
the kinetic energy of the relativistic jet into gamma-ray
emission is $\eta=E_{\gamma,iso}/(E_{\gamma,iso}+E_{K,iso})\sim 0.4$
(with a large scatter of individual sources) (Fong
et al. 2015). This allows the kinetic energy of the jet
to be used to estimate the possible power of the short
GRB produced by it.

To be specific, consider the Blandford–Znajek process as a possible physical mechanism
of a short GRB (Nakar 2007). The energy of the
BZ jet is determined by the magnetic field around
the BH and its spin, $L_{BZ}\sim \Phi^2\Omega_H^{*2} f(\Omega_H^*)$, where  $\Phi$ is the magnetic flux through the BH ergosphere, $\Omega_H^*=(1/2)a^*/(1+\sqrt{1-a^{*2}})$ 
is the dimensionless angular velocity of rotation on the BH horizon, and $f(\Omega_H^*)\approx 1+1.38 \Omega_H^{*2}-9.2\Omega_H^{*4}$ is the correction
function that can be derived by fitting the numerical
calculations (Nakar 2007). The magnetic field is
an uncertain parameter, but it can be eliminated by
assuming the balance between the magnetic and
dynamic pressures in the disk around the BH. In
this case, $L_{BZ}\sim \dot M c^2\Omega_H^{*2} f(\Omega_H^*)$.
Assuming the accretion rate to be $\dot M=M_d/\Delta t$, where $\Delta t$
is the accretion time, the kinetic energy of the BZ
jet is found to be $\Delta E_{K,iso}\approx 0.015 M_dc^2\Omega_H^2f(\Omega_H)$
(we use the numerical coefficient justified in Barbieri
et al. 2019).

Figure 5 presents the cumulative distributions of
the kinetic energy of the BZ jet $\Delta E_{K,iso}$ for coalescing
NS+BH binary systems within the detection
horizon $D_h (O3)$ at the current phase of O3 observations
with the LIGO/Virgo GW interferometers with
allowance made for the evolution of the metallicity $Z$
and the mean star formation rate $SFR$ in galaxies for
two common envelope parameters. We see a strong
dependence on the NS equation of state (more energetic
jets are obtained at greater values of the tidal
deformability  $\Lambda$) and on the degree of approach of the
binary components at the common envelope stage.
Less efficient common envelopes (a large parameter
 $\alpha_{CE}$) lead to a noticeably larger percentage events with astrophysically interesting EM energy release

\begin{figure*}
	\includegraphics[width = \columnwidth]{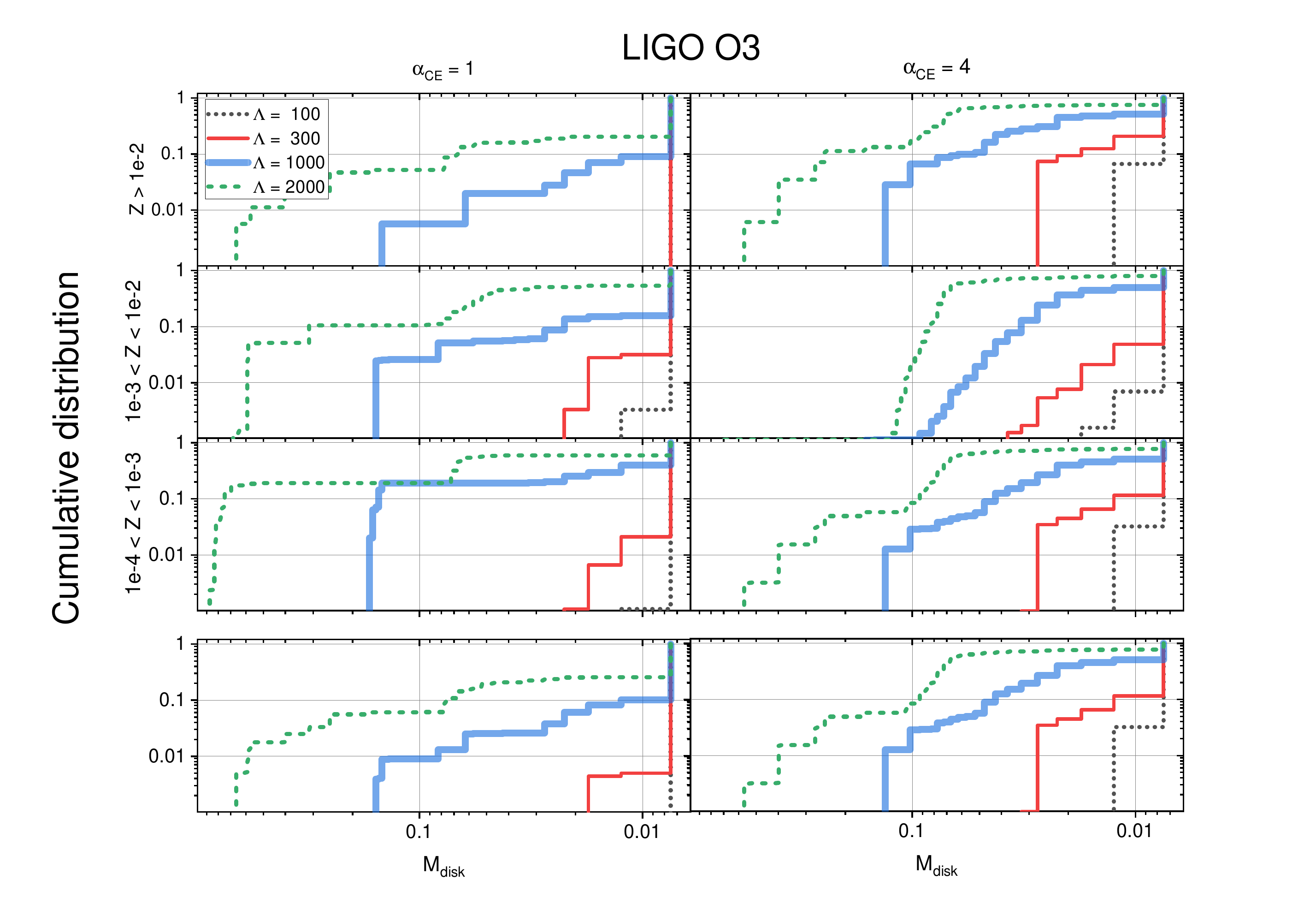}
	\caption{(Color online) Cumulative distributions of the mass of the remnant disk around the BH after NS tidal disruption. The
color lines in the inset indicate the NS tidal deformability $\Lambda$ parameterizing the various NS equations of state. The upper
three rows present the results for various stellar metallicities. The fourth row presents the result with allowance made for the
time evolution of the mean star formation rate and the stellar metallicity in galaxies. The left and right columns present the calculations for the common envelope parameters $\alpha_{CE}=1$ and $\alpha_{CE}=4$, respectively.}
	\label{fig:mdO3}
\end{figure*}
\begin{figure*}
	\includegraphics[width= \columnwidth]{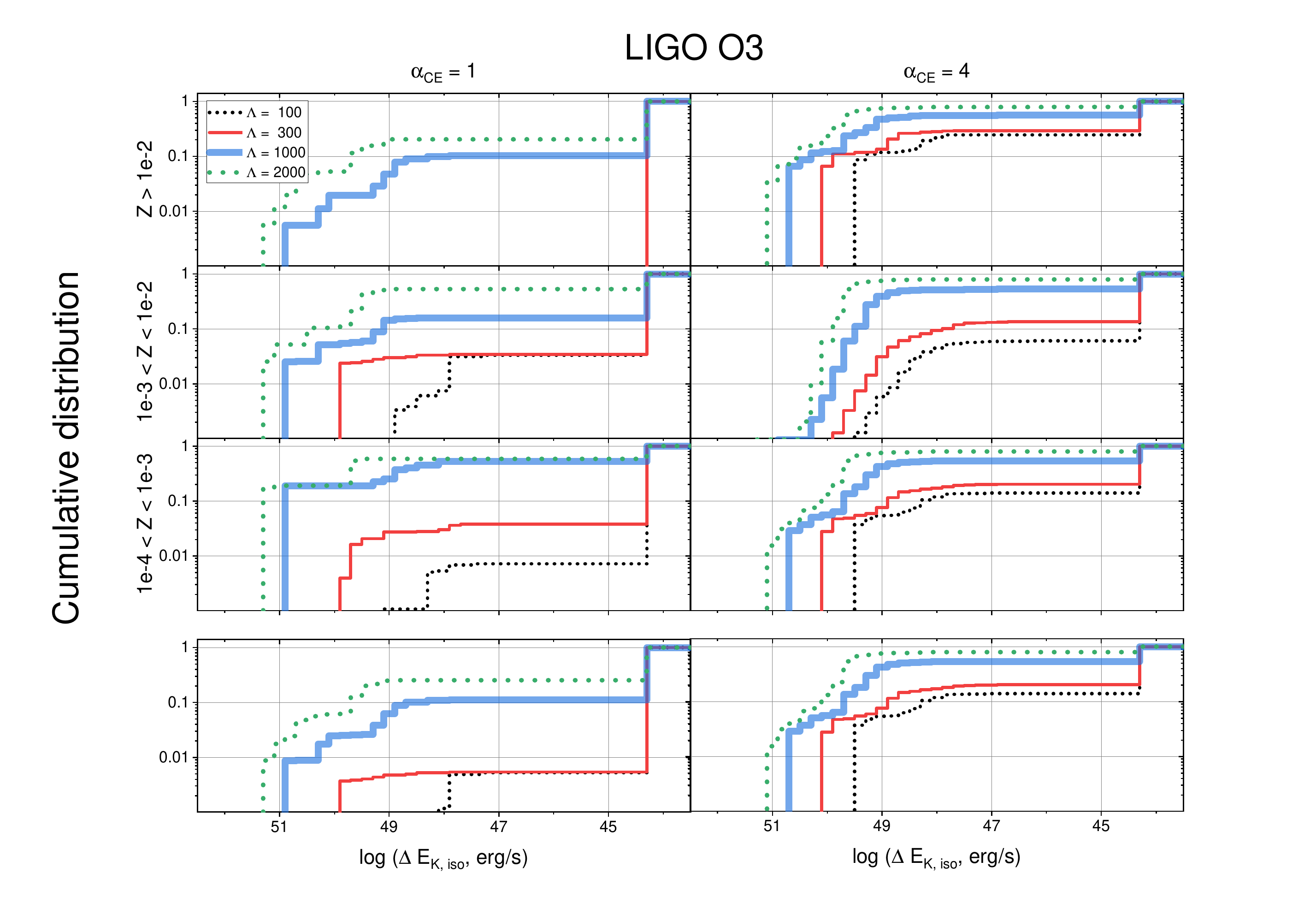}
	\caption{
	(Color online) Same as Fig. 4, but for the kinetic energy of the jet launched by the Blandford–Znajek mechanism from
the remnant disk around the rotating BH after the coalescence of NS+BH binaries.}
	\label{fig:EO3}
\end{figure*}

\section{COALESCENCES WITHOUT NS TIDAL
DISRUPTION}
\label{s:plunge}

The NS+BH coalescences in binaries with a
large mass ratio $q$ occurring without tidal disruption
can also be of interest from the viewpoint of the appearance
of accompanying EM radiation. The NS
should have a magnetic field, with the NS+BH coalescences
occurring in a fairly short time after the formation
in most cases, so that the NS magnetic field
has no time to decay. Several physical mechanisms
for the generation of EM radiation associated with
electrodynamic processes in the vicinity of a BH coalescing
with a magnetized NS (see, e.g., Zhang 2016;
Levin et al. 2018; Zhang 2019; Dai 2019; and references
therein) or with the fundamental graviton-to-photon
conversion in a magnetic field (Dolgov and
Postnov 2017) are possible in this case.

\subsection{Electric Charge of a Rotating BH}

The spin and orbital motion of an electrically
charged BH in a coalescing binary system initiate
time-varying electric dipole and magnetic dipole moments
in the binary that give rise to EM radiation (for
the estimates and discussion, see Dai 2019), whose
power and energetics for rapidly rotating BHs can be
sufficient for the explanation of short EM transients
(for example, fast radio bursts (FRBs); see Popov
et al. 2018)). A rotating BH in an external magnetic
field can acquire an electric charge with a maximum
value of$Q_w=(2G/c^3)JB$, where $J=a^*GM_{BH}^2/c$
is the angular momentum of the rotating BH and
$B$ is the magnetic field strength (Wald 1974). This
mechanism has also been recently discussed for the
estimates of a possible EM radiation pulse already
after the NS+BH coalescence (Zhong et al. 2019).
It is convenient to normalize the charge of a rotating
Reissner–Nordstr\"om BH to the characteristic value
of  $Q_{RN}=2\sqrt{G} M_{BH}\approx 10^{30}(M_{BH}/M_\odot)$  (emu) corresponding
to the equality of the Schwarzschild
radius to the Reissner–Nordstr\"om radius: 
$\tilde q_w=Q_W/Q_{RN}$.

\begin{figure*}
	\includegraphics[width= \columnwidth]{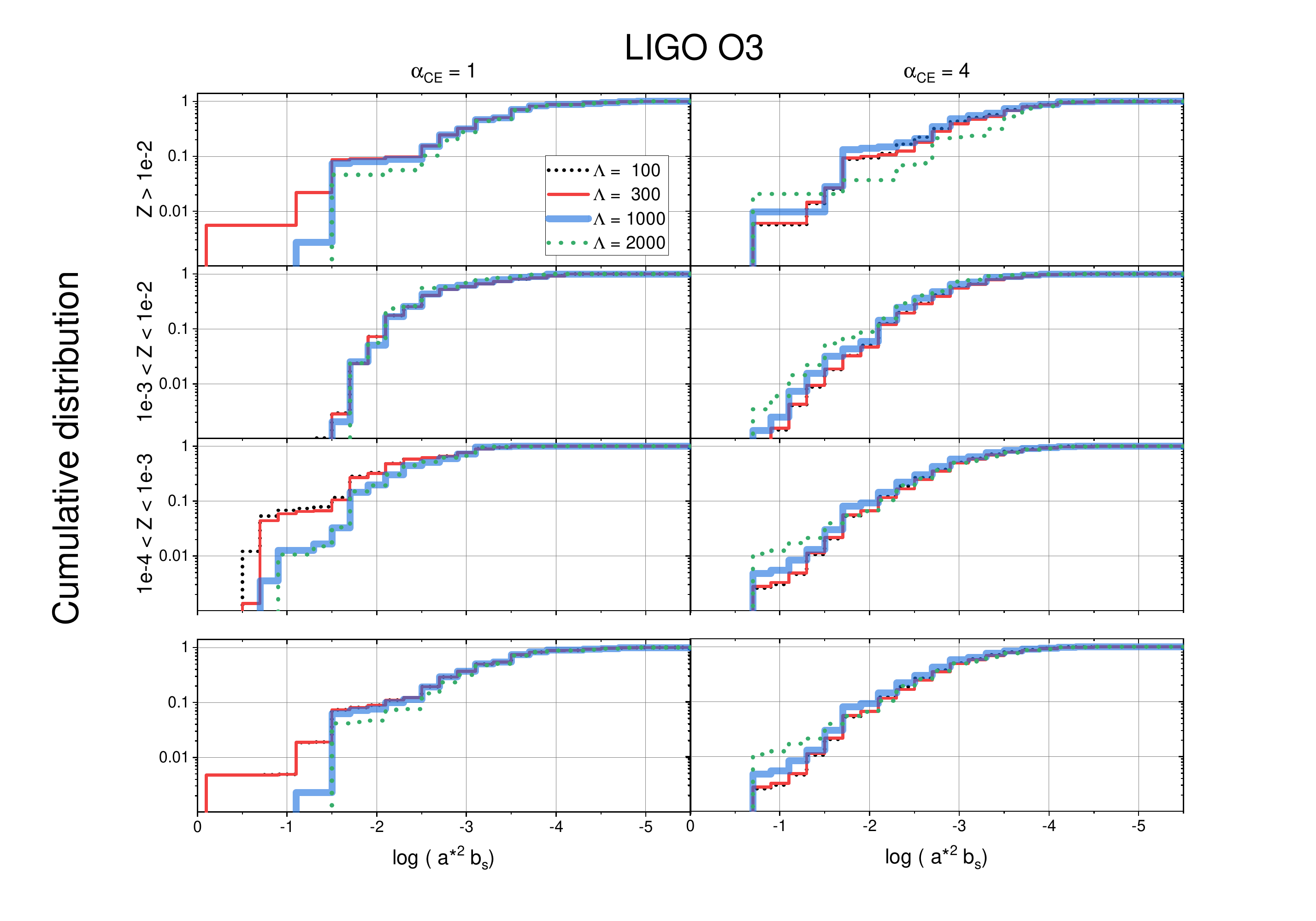}
	\caption{(Color online) Same as Fig. 4, but for the combination of the BH spin before the coalescence and the NS magnetic field $a^{*2}b_s$ determining the maximum intrinsic magnetic dipole moment of aWald-charged BH $\mu_{W,max}$ (Eq. (2)).}	\label{fig:muO3}
\end{figure*}
\begin{figure*}
	\includegraphics[width= \columnwidth]{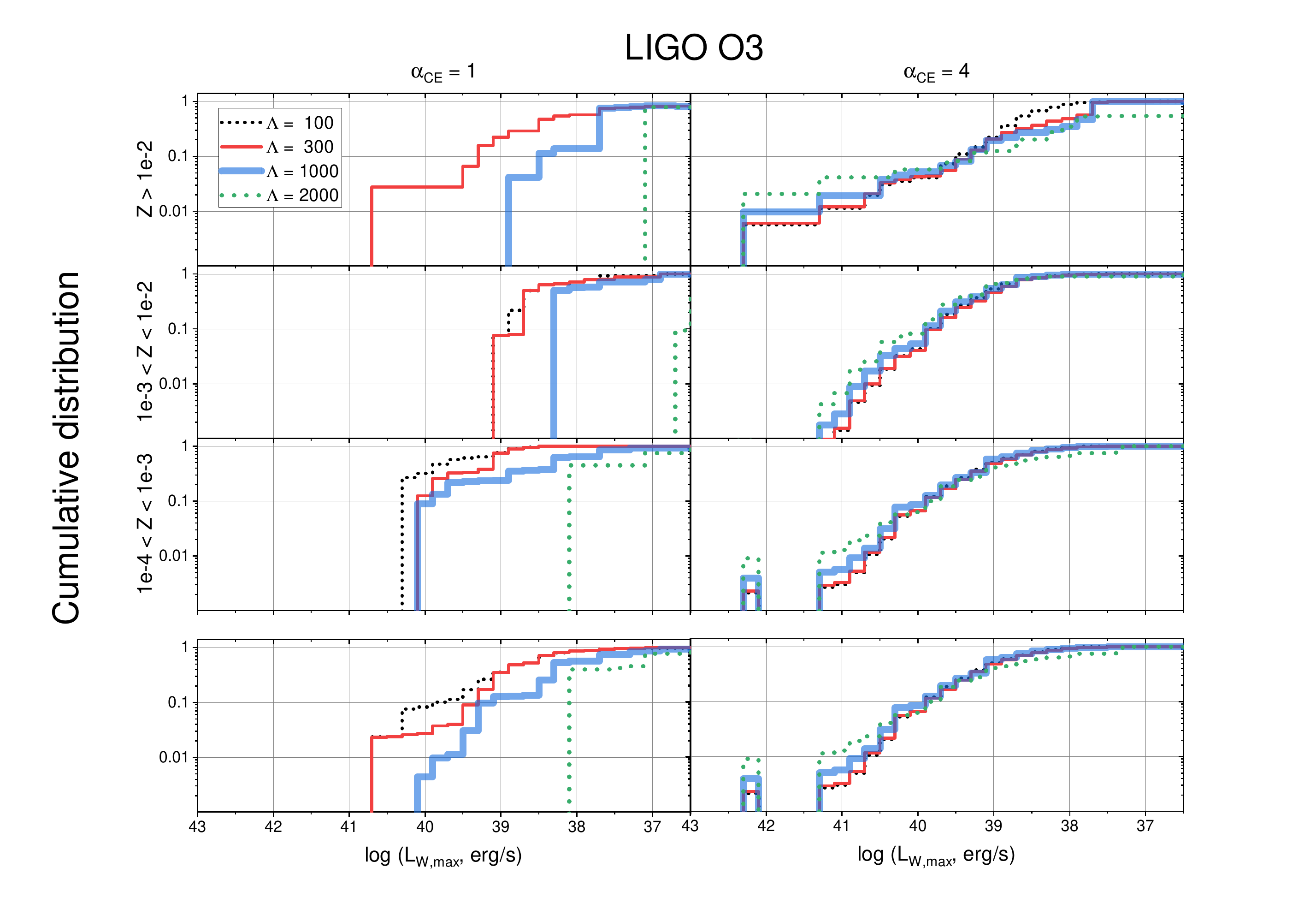}
	\caption{(Color online) Same as Fig. 4, but for the maximum energy release from a charged rotating BH after the coalescence $L_{W,max}$ (Eq. (3)).}
	\label{fig:LWO3}
\end{figure*}

In natural units $\hbar=c=1$ Newton’s gravitational
constant is written via the Planck mass 
$G=1/m_{Pl}^2$, $m_{Pl}\approx 10^{19}$~GeV, the electric charge is dimensionless,
the electron charge is expressed via the fine structure
constant $\alpha = 1/137$ as $e^2=4\pi \alpha$. It is also
convenient to make the magnetic field dimensionless
by normalizing it to the critical (Schwinger) one
$b=B/B_{cr}$, where $B_{cr}=m_e^2/e\approx 4.41\times 10^{13}$~G ($m_e\approx 511$~keV
is the electron rest mass). The specific Wald
BH charge is then
\beq{e:qw}
\tilde q_W=\frac{a^*b}{\sqrt{4\pi\alpha}}\myfrac{m_e}{m_{Pl}}^2\myfrac{M_{BH}}{m_{Pl}}\approx 10^{-6}a^*b\myfrac{M_{BH}}{M_\odot}\,.
\eeq
This formula is written for a uniform magnetic field.
Note that for a NS dipole field $b(R)=b_s(R_{NS}/R)^3$,
where $b_s$ is the surface field, it follows from (1) that
at the tidal radius $R_t\sim R_{NS}q^{1/3}$ the specific Wald charge does not depend on the BH mass:  $\tilde q_W\sim 10^{-6}a^*b_s(M_{NS}/M_\odot)$.

For the electric dipole and magnetic dipole radiation
associated with the orbital motion of a charged
BH before the coalescence, the EM radiation power is
proportional to the square of the NS magnetic field $b$
and the square of the BH spin, $\sim a^{*2} b^2$. In the case of magnetic dipole radiation from the rotating charged
BH itself acquiring the Wald charge at the stage
when the charged NS approaches the BH before the
coalescence, it is proportional to the square of the
intrinsic magnetic moment 
$\mu_W^2\sim a^{*4} b^2$ (Dai 2019).
Therefore, the Wald charge can be important only
for rapidly rotating BHs with $a^*\gtrsim 0.5$. The number
of such binaries before the coalescence is extremely
small (see the dashed curve in Fig. 3).

If the NS is swallowed by the BH without being disrupted,
the maximum Wald charge can be estimated
from the magnetic field at the Schwarzschild
BH radius $R\sim R_g=2GM_{BH}/c^2$ before the coalescence:
$\tilde q_{W,max}\sim 3\times 10^{-5} a^*b_s (R_{NS}/10\hbox{km})^3(M_{BH}/M_\odot)^{-2}$.
In this case, the maximum intrinsic magnetic dipole
moment of the charged BH will be determined only
by the BH spin $a^*$, NS magnetic field $b_s$, and NS
radius:
\beq{e:mumax}
\mu_{W,max}=\frac{J_{BH}Q_{W,max}}{M_{BH}c}\approx 5\times 10^{30}
 a^{*2}b_s\myfrac{R_{NS}}{10\hbox{km}}^3\,.
\eeq

The maximum energy release from a charged rotating
single BH with a magnetic moment $\mu_{W,max}$ can
be estimated from the magnetic dipole formula
$L_{W,max}\sim \mu_{W,max}^2\Omega_H^4$, where  $\Omega_H^f\sim (a^{*}_f/2)(c^3/GM_{BH}^f)$ is the angular velocity of the horizon for a BH with mass $M_{BH}^f$
BH after the
coalescence. Substituting $\mu_{W,max}$ from (2), we find
\beq{e:Lwmax}
L_{W,max}\sim  10^{42}[\hbox{erg/s}]
 a^{*4}b_s^2a^{*4}_f\frac{(R_{NS}/10\hbox{km})^6}{(M_{BH}^f/10 M_\odot)^4}\,.
\eeq
This estimate is comparable in magnitude to the
estimates of the possible EM energy release from
charged BHs at the pre-coalescence (in-spiraling)
stage (Zhang 2019; Dai 2019). Given the relative
smallness of the BH spins before ($a^*\sim 0.2$, the
dashed curve in Fig. 3) and after ($a^{*,f}\sim 0.6$
the solid curve in Fig. 3) the coalescence as well as the
low (poorly known) Poynting flux-to-EM radiation
conversion efficiency, there can be astrophysically
interesting EM energy release in this process only
for rapidly rotating low-mass BHs in a pair with
strongly magnetized NSs with $b_s\sim 1$, whose number
is extremely small.


\subsection{Conversion of Gravitational Waves in a Relativistic
Plasma in a Magnetic Field}

At the final stages before the coalescence of a binary
magnetized NS+BH, conditions for the additional
appearance of EM radiation due to the coherent conversion
of gravitational waves into electromagnetic
ones in a magnetic field arise in the binary. For
a vacuum this mechanism was first considered by
Gertsenshtein (1962). In the presence of a surrounding
plasma the effect in cosmological applications
was considered by Dolgov and Ejlli (2012), while for
the case of conversion in a nonrelativistic plasma with
a magnetic field around astrophysical GW sources,
coalescing binary NSs and BHs, it was considered
by Dolgov and Postnov (2017). In the latter paper
it was emphasized that since the plasma frequency in
the interstellar medium with density $n_e\sim 1$~cm$^{-3}$, $\Omega_e=60\ \sqrt{n_e}$ kHz, is much greater than the frequency
of the GWs from coalescing binary systems (100–
200 Hz), then the GW-to-EM conversion in the
plasma is dissipative in nature and is determined
by the imaginary part of the dielectric permittivity.
The EM wave damping amplitude $A_j$ was shown to
be related to the amplitude $h_j$ of a GW propagating
with frequency $\omega$ perpendicularly to an external magnetic
field B by the relation (in natural units $\hbar=c=k_B=1$)
\begin{equation}
\label{eq:10}
A_j\approx \frac{\kappa b\omega a_e}{\Omega_e} h_j\,,
\end{equation}
where $ a_e = \sqrt {\frac {T_e} {e^2 n_e}} $ 
is the Debye radius of electrons, $T_e$ is the electron temperature, $\kappa^2=16\pi/m_{Pl}^2$ is the coupling constant. 

The fraction of the energy being
released in the GW that dissipates into the thermal
energy of the plasma,
\beq{}
\label{eq:11}
K_{nr} 
= {\left(\frac{\kappa B\omega a_e}{\Omega_e}\right)}^2\approx 10 ^ {- 46} \left (\frac {\omega} {\Omega_e} \right) ^ 2 \left (\frac {a_e} {\hbox{cm}} \right) ^ 2 \left ( \frac {B} {1 \ \hbox{G}} \right)^2
\eeq
is very small and is interesting only for superstrong
magnetic fields. It may well be that part of this
energy can be reprocessed into high-frequency radio
emission (Marklund et al. 2000).

During the coalescence of magnetized NSs with
BHs, the magnetosphere of a NS with a magnetic
field $B_{NS}$ spinning with a frequency $\omega_{NS}=2\pi/P_{NS}$
is filled with an ultrarelativistic plasma with a density
no less than the Goldreich–Julian charge density
$n_{GJ}=(\omega_{NS}B_{NS})/(2\pi e)$. For radio pulsars 
$n_e=\lambda n_{GJ}$, where $\lambda\sim 10^4-10^5$
is the multiplicity factor
of the pairs created near the NS surface (Beskin
2018). In a relativistic collisional plasma with
a plasma frequency 
 $\Omega_{rel}^2=\frac{4 \pi e^2 n_e}{3 T_e}$, $T_e\sim \gamma m_e$ ($\gamma$ is the
electron Lorentz factor) we can obtain (Postnov and
Simkin 2019)
\begin{equation}
\label{eq:17}
A_j \approx   \frac{\kappa b\omega }{\Omega^2_{rel}}h_j\sim\frac{\kappa b\omega}{(n_e/\gamma m_e)}\,.
\end{equation}

If the plasma flows along open magnetic field
lines (as in a pulsar), then for a dipole field $B(R)\sim B_s(R_{NS}/R)^3$. In a magnetic flux tube the magnetic
flux is conserved, $\Phi=B(R)S(R)=const$; given the
continuity equation $n_eB(R)S(R)=const$ , we then
find that for an ultrarelativistic collisional plasma with
$T_e\sim \gamma m_e$ the conversion efficiency does not depend on the NS magnetic field, but is determined only by
the lepton Lorentz factor, the NS spin period, and the
pair multiplicity with respect to the Goldreich–Julian
density $\lambda$:
\begin{eqnarray}
\label{eq:25}
&K_{rel}=\left(\frac{\kappa b\omega}{\Omega^2_{rel}}\right)^2 \nonumber \\
&\approx 10^{-35}\left(\frac{\omega }{100\ \hbox{rad}\, \hbox{s}^{-1}}\right)^2\left(\frac{P_{NS}}{1\ \hbox{s}}\right)^2\left(\frac{\lambda }{10^5}\right)^{-2}\left(\frac{\gamma }{10^5}\right)^2\,.
\end{eqnarray}

Clearly, the effect is weak for the densities
of a relativistic pulsar plasma with $\lambda\sim 10^4-10^5$.
However, the relativistic plasma density near a coalescing
NS+BH binary is unknown and, therefore,
the density $n_{GJ}$ can serve only as an approximate
lower limit. The upper bound of the GW-to-EM conversion efficiency in a relativistic plasma
around a coalescing NS+BH binary will then be
$K_{rel}\lesssim 10^{-25}(P_{NS}/1\hbox{s})^2(\gamma/10^5)^2(n_e/n_{GJ})^{-2}$, i.e.,
for a GW pulse energy during the coalescence of 
$M_\odot c^2\sim 2\times 10^{54}$ erg, up to $\sim 10^{38}$ ergs can be additionally
reprocessed into the thermal energy of the
plasma.

\section{CONCLUSION}
\label{s:conclusion}

In this paper we analyzed the various physical
mechanisms that could give rise to an EM pulse
accompanying the coalescence of magnetized NS+BH binaries. At the time of writing this paper (mid-
October 2019), the LIGO/Virgo detectors recorded
several such binaries, but no EM radiation from them
have been detected so far. Based on a series of
population synthesis calculations (Postnov and Kuranov
2019; Postnov et al. 2019), we constructed
the distributions of the NS+BH binary coalescence
rate density and the expected detection rate of such
binaries in the current LIGO/Virgo O3 observations
by taking into account the evolution of the stellar
metallicity and the star formation rate in the Universe
(Fig. 1).

The derived distributions of coalescing NS + BH
in component mass ratio $q=M_{BH}/M_{NS}>1$, magnetic fields $b_s=B_{NS}/B_{cr}$, $B_{cr}=4.14\times 10^{13}$~ G,
NS spin period, and BH angular momentum (dimensionless
spin $a^*=J_{BJ}/(GM^2_{BH}/c)$) before the coalescence
were used to estimate the mass of the remnant
disk around the BHMd (Fig. 4) and the BH spin
after the coalescence $a^*_f$
(Fig. 3). We estimated the
masses of the remnant baryonic disks, the BH masses
and spins after the coalescence based on interpolation
of the relativistic numerical calculations by Zappa
et al. (2019) and Jimenez-Forteza et al. (2017) by
taking into account the NS equation of state that was
specified by the dimensionless tidal deformability $\Lambda$.
Assuming that the magnetic field in the remnant disk
around a rotating BH to be dynamically balanced,
we estimated the kinetic energy of the relativistic
jet launched by the BZ mechanism (Fig. 5), which
depends strongly on the NS equation of state through the tidal deformability $\Lambda$.

We separately studied the NS+BH coalescences
with a large mass ratio $q$ occurring without NS tidal
disruption. For such binaries we constructed the
distributions of the maximum possible BH magnetic
dipole moment before the coalescence $\mu_{W,max}$ 
acquired due to the Wald BH charge (Wald 1974;
Levin et al. 2018) (Fig. 6) and corresponding to
the maximum EM magnetic dipole luminosity of
such a charged BH $L_{W,max}$ (Fig. 7). Even for the
most optimistic parameters our estimates of the EM
luminosity from a Wald-charged BH in coalescing
NS+BH binaries are considerably smaller than those expected from binaries with a smaller mass
ratio, where the formation of remnant disks and
relativistic jets is possible.

We additionally considered the conversion of gravitational
waves in a magnetic field with a relativistic
plasma that may surround a NS+BH binary at the
pre-coalescence stage. This mechanism was shown
to convert no more than $\sim 10^{38}-10^{39}$ erg into additional plasma heating even under the most favorable
conditions (a large lepton Lorentz factor  $\gamma \sim 10^5$ and a low plasma density of the order of the Goldreich-Julian one).

Our general conclusion is that noticeable EM
phenomena from coalescing NS+BH binaries might
be expected from a small fraction of the coalescences in which the NS is tidally disrupted and a remnant disk is formed around the rotating BH. The fraction of
such events depends on the mass ratio $q$ and the NS
equation of state. At the expected coalescence rate
of several events per year in the current LIGO/Virgo
O3 GW observations, the chances to see a weak EM
signal are slim. The more exotic physical mechanisms
(the Wald electric charge of a rotating BH or
the fundamental conversion of gravitational waves
into electromagnetic ones in a magnetized plasma
surrounding a coalescing NS+BH binary) are much
less efficient. At the current detector sensitivity level
EM phenomena from coalescing NS+BH binaries
in various ranges might be expected only from the
nearest events (at distances of tens of Mpc). The
detection of an EM counterpart from NS+BH can
potentially give physically rich information about the equation of state and the magnetic field of neutron stars.


\textbf{Acknowledgements.}
The work of K.A. Postnov was supported by RSF
grant no. 19-12-00229. The work of A.G. Kuranov was
supported by the Scientific School of the Moscow State University “Physics of Stars, Relativistic Objects, and Galaxies.”

\section*{References}
1.	B. P. Abbott, R. Abbott, T. D. Abbott, et al., Phys. Rev. Lett. 116, 241103 (2016a).

2.	B. P. Abbott, R. Abbott, T. D. Abbott, et al., Phys. Rev. Lett. 116, 061102 (2016b).

3.	B. P. Abbott, R. Abbott, T. D. Abbott, LIGO Sci. Collab., and Virgo Collab., Phys. Rev. X 9, 011001 (2019).

4.	C. Barbieri, O. S. Salafia, A. Perego, M. Colpi, and
G.	Ghirlanda, arXiv:1908.08822 (2019).

5.	V. S. Beskin, Phys. Usp. 61, 353 (2018).

6.	R. D. Blandford and R. L. Znajek, Mon. Not. R. Astron. Soc. 179, 433 (1977).

7.	Z. G. Dai, Astrophys. J. Lett. 873, L13 (2019).

8.	T. Damour, A. Nagar, and L. Villain, Phys. Rev. D 85, 123007 (2012).

9.	A. D. Dolgov and D. Ejlli, J. Cosmol. Astropart. Phys. 2012 (12), 003 (2012).

10.	A. Dolgov and K. Postnov, J. Cosmol. Astropart. Phys. 2017 (9), 018 (2017).

11.	W. Fong, E. Berger, R. Margutti, and B. A. Zauderer, Astrophys. J. 815, 102 (2015).

12.	M. Gertsenshtein, Sov. Phys. JETP 14, 84 (1962).

13.	R. Gill, A. Nathanail, and L. Rezzolla, Astrophys.J.	876, 139 (2019).

14.	J.-L.Jiang,S.-P.Tang,D.-S.Shao,et al., arXiv:1909.06944 (2019).

15.	X. Jim enez-Forteza, D. Keitel, S. Husa, et al., Phys. Rev. D 95, 064024 (2017).

16.	K. Kawaguchi, K. Kyutoku, M. Shibata, and
M.	Tanaka, Astrophys. J. 825, 52 (2016).

17.	K. Kyutoku, K. Ioka, H. Okawa, M. Shibata, and K.	Taniguchi, Phys. Rev. D 92, 044028 (2015).

18.	J. Levin, D. J. D’Orazio, and S. Garcia-Saenz, Phys. Rev. D 98, 123002 (2018).

19.	LIGO/Virgo Sci. Collab., arXiv e-prints (2018).

20.	The LIGO Sci. Collab., the Virgo Collab., J. Abadie, et al., arXiv:1003.2481 (2010).

21.	The LIGO Sci. Collab., the Virgo Collab., et al., arXiv:1908.01012 (2019).

22.	V. M. Lipunov, K. A. Postnov, M. E. Prokhorov, and E. Y. Osminkin, Astrophys. J. 423, L121 (1994).

23.	M.Marklund, G. Brodin,and P.K.S. Dunsby, Astrophys. J. 536, 875 (2000).

24.	B. D. Metzger, arXiv:1910.01617 (2019).

25.	E. Nakar, Phys. Rep. 442, 166 (2007).

26.	G. Nelemans and C. A. Tout, Mon. Not. R. Astron. Soc. 356, 753 (2005).

27. E. Pfahl, P. Podsiadlowski, and S. Rappaport, Astrophys. J. 628, 343 (2005).

28.	L. Piro, E. Troja, B. Zhang, et al., Mon. Not. R. Astron. Soc. 483, 1912 (2019).

29.	S. B. Popov, K. A. Postnov, and M. S. Pshirkov, Phys. Usp. 61, 965 (2018).

30. K. Postnov, A. Kuranov, and N. Mitichkin, Phys. Usp. 62, 1153, (2019); arXiv:1907.04218 (2019).

31. K. A. Postnov and A. G. Kuranov, Mon. Not. R. Astron. Soc. 483, 3288 (2019).

32. K. A. Postnov and I. V. Simkin, Journal of Physics: Conf. Ser. 1390, 012086 (2019).

33. K. A. Postnov and L. R. Yungelson, Liv. Rev. Relativ. 17, 3 (2014).

34.	K. A. Postnov, A. G. Kuranov, D. A. Kolesnikov, S. B. Popov, and N. K. Porayko, Mon. Not. R. Astron. Soc. 463, 1642 (2016).

35.	M. Shibata and K. Hotokezaka, Ann. Rev. Nucl. Part. Sci. 69, (2019).

36.	M. Shibata and K. Taniguchi, Liv. Rev. Relativ. 14,6 (2011).

37.	A. M. Soderberg, S. R. Kulkarni, E. Nakar, et al., Nature (London, U.K.) 442 (7106), 1014 (2006).

38.	A. Tchekhovskoy, R. Narayan, and J. C. McKinney, Astrophys. J. 711, 50 (2010).

39.	R. M. Wald, Phys. Rev. D 10, 1680 (1974).

40.	L. R. Yungelson, A. G. Kuranov, and K. A. Postnov, Mon. Not. R. Astron. Soc. 485, 851 (2019).

41.	F. Zappa, S. Bernuzzi, F. Pannarale, M. Mapelli, and N.	Giacobbo, Phys. Rev. Lett. 123, 041102 (2019).

42.	B. Zhang, Astrophys. J. Lett. 827, L31 (2016).

43.	B. Zhang, Astrophys. J. Lett. 873, L9 (2019).

44.	S.-Q. Zhong, Z.-G. Dai, and C.-M. Deng, arXiv:1909.00494 (2019).
\label{lastpage}

\end{document}